**Distinguishing Inner and Outer-Sphere Hot Electron Transfer in Au/p-GaN Photocathodes**

*Fatemeh Kiani[1], Alan R. Bowman[1], Milad Sabzehparvar[1], Ravishankar Sundararaman[2], Giulia Tagliabue[1]\**

[1] *Laboratory of Nanoscience for Energy Technologies (LNET), STI, École Polytechnique Fédérale de Lausanne, 1015 Lausanne, Switzerland*

[2] *Department of Materials Science & Engineering, Rensselaer Polytechnic Institute, 110 8th Street, Troy, New York 12180, USA*

*\*E-mail: giulia.tagliabue@epfl.ch*

**Abstract**

Exploring nonequilibrium hot carriers from plasmonic metal nanostructures is a dynamic field in optoelectronics, driving photochemical reactions such as solar fuel generation. The hot carrier injection mechanism and the reaction rate are highly impacted by the metal/molecule interaction. However, determining the primary type of the reaction and thus the injection mechanism of the hot carriers has remained elusive. In this work, we reveal an electron injection mechanism deviating from a purely outer-sphere process for the reduction of ferricyanide redox molecule in a gold/p-type gallium nitride (Au/p-GaN) photocathode system. Combining our experimental approach with ab initio simulations, we discover that the efficient inner-sphere transfer of low-energy electrons leads to a continuous enhancement in the photocathode device performance in the interband regime. These findings provide important mechanistic insights, showing our methodology as a powerful tool for analyzing and engineering hot-carrier-driven processes in plasmonic photocatalytic systems and optoelectronic devices.

**Introduction**

Plasmonic catalysis shows great promise in improving reaction rates and selectivity across various catalytic applications.[1–5] Three potential mechanisms can occur to drive reactions: enhanced electromagnetic near fields, temperature increase at the metal/liquid interface, and transfer of excited hot carriers.[6–9] In particular, the last of these processes has attracted attention for the possibility of steering chemical reactions by accessing highly excited unoccupied states.[5,10–12] Practical applications of hot carrier devices requires a full understanding of plasmonic hot-carrier-driven processes including plasmon excitation, hot carrier generation, and injection at interfaces. However, the microscale physical understanding of hot carrier transport and injection mechanisms at the metal/molecule and metal/semiconductor interfaces in these devices remained elusive.

Hot carrier collection schemes typically involve the formation of an interfacial Schottky barrier between plasmonic metals (e.g., Au) and wide band gap semiconductors (e.g., $TiO_2$).[8] This strategy enables us to quickly capture hot electrons (hot holes) into the conduction band of an n-type (p-type) semiconductor and selectively harvest non-equilibrium hot holes (hot electrons) from the metal nanostructure by driving an oxidation (reduction) reaction. It has been recently shown that the energy of distribution of the hot carriers and their ballistic transport to the metal/semiconductor or metal/molecule interface play a critical role for the internal quantum efficiency of plasmonic hot carrier devices.[8,13,14] In addition to the energy distribution of hot carriers, the interaction of molecules with the metal surface, and thus the injection mechanism of the hot carriers to the molecule, will play a role, leading to expected differences between outer- and inner-sphere reactions. In an outer-sphere reaction, the hot carrier transfer between the metal and the molecule occurs at a plane separated by a solvent layer from the metal surface (**Figure 1.a**).[15] This introduces a tunneling barrier for the carrier transfer to the molecule.[16] Conversely, in an inner-sphere reaction the carrier transfer can take place through the bridging ligand of the adsorbed molecule bonds to the metal surface.[15,17] Thus, energy-dependent hot carrier injection process and the rate of the reaction are expected to be highly impacted by the metal /molecule interaction.[5,12,18–20] To date, determining the reaction type and thus the injection mechanism of the hot carriers remains challenging.

In this work, we study plasmonic Au/p-GaN photocathodes to drive reduction of ferricyanide redox molecule, $Fe(CN)_6^{3-}$ whilst using a p-type semiconductor to collect the hot holes, reducing charge recombination. Leveraging scanning electrochemical microscopy (SECM) and highly controlled monocrystalline, ultrathin (14-18 nm) gold nanodisk antennas, we quantify the internal quantum efficiency (IQE) of our devices and from it determine the energy-resolved injection probability of hot electrons. Our study reveals two coexisting mechanisms: an outer-sphere transfer of high-energy electrons that inject into the $Fe(CN)_6^{3-}$ molecule through a tunneling process, and an inner-sphere transfer of low-energy electrons that directly inject into the LUMO of the molecule (**Figure 1.a**). We suggest the latter comes from the higher affinity of the $Fe(CN)_6^{3-}$ molecules to adsorb on the surface, and results in an IQE growth in the interband regime for our photocathode device. Moreover, we observe a ballistic collection of high-energy d-band holes at the Au/p-GaN interface. Overall, this comprehensive mechanistic understanding highlights the robustness of our experimental and theoretical approach for analyzing charge transport processes at interfaces in plasmonic hot-carrier-driven photodetection and photocatalytic systems. Furthermore, considering that a similar photocathodes was previously used for photoelectrochemical $CO_2$ reduction,[3,4] our results contribute to the directed development of plasmonic photocathodes for artificial photosynthesis.

**Results and discussions**

We fabricated plasmonic photocathodes consisting of an array of single-crystalline Au nanodisks (SC Au NDs) with sizes of the order of tens of nanometers on an optically transparent p-type GaN (p-GaN) on a sapphire substrate (**Figure 1.b**). The nanoantennas were fabricated from single-crystalline Au microflakes (SC Au MFs)[21] with varying thicknesses of 14-27 nm (see Methods in **Supplementary Information 1.1**) to study the effect of nanoantenna thickness on hot-carrier-driven processes within our device. We use p-GaN as the supporting semiconductor because of its large optical band gap ($E_G \simeq 3.4$ eV)[22] that prevents visible-light absorption within the p-GaN film, and the suitable Schottky barrier ($\Phi_B$) across the Au/p-GaN heterojunction (**Figure S4.b**), which enables selective hot-hole collection (**Figure 1.b**). The Au NDs are in contact with an electrolyte containing a reversible redox molecule, $Fe(CN)_6^{3-}$ (ferricyanide, the oxidized form, *Ox*) which enables hot-electron collection through a photochemical reduction reaction (**Figure 1**). The reduction of this molecule is expected to proceed via a one-electron transfer, non-purely outer-sphere mechanism with fast kinetics.[17,23–25] The chosen molecule also does not absorb visible light (**Figure S1**, 470-832 nm).

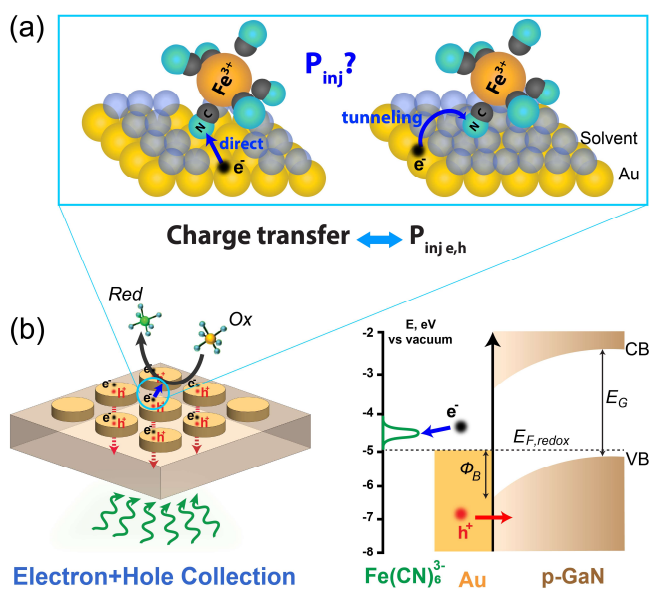

**Figure 1.** Schematic of interfacial hot carrier collection in plasmonic metal/semiconductor heterostructure devices. (a) Injection mechanisms of hot electrons across the metal/molecule interface. (b) Au NDs/p-GaN photocathode in contact with an oxidant (*Ox*) molecule, together with the band alignment showing hot-hole and hot-electron collection across the Au/p-GaN and $Fe(CN)_6^{3-}$ interfaces.

For photochemical experiments, we performed a series of scanning photo-electrochemical microscopy (photo-SECM)[8] measurements on Au ND arrays with disk thicknesses of 14, 16, and 18 nm, and average diameters of 52, 68, and 67 nm, respectively. We present the implemented photo-SECM approach in **Figure 2.a**. We performed photo-SECM measurements in substrate generation/tip collection (SG/TC) mode, where a reduction reaction proceeds at the substrate, and the generated reductant species get oxidized at the UME tip (**Figure 2.a**). During the SG/TC SECM experiment, we applied an oxidation potential of 0.4 V vs Ag/AgCl at the Pt UME tip and the substrate was kept at open circuit condition to just study the photo effect. We illuminated the substrate from the bottom (see Methods in **Supplementary Information 1.2**) to drive the photo-reduction reaction only at the illuminated area of the substrate upon hot-carrier generation and hot-electron transfer at the Au/electrolyte interface. **Figure S5.a** shows the time-trace of the tip current ($i_{Tip}$) upon illumination of the 14 nm thick ND array (**Figure 2.a**, SEM image) at a broad excitation wavelength range of 470-832 nm, where the intensity was modulated up to 14 W/cm$^2$. At this power intensity range, we showed no local heating effect. This is supported by both experimental and theoretical confirmation (see our previous work).[8] The measured tip current relative to the dark current ($i_{Tip}/i_{Tip,dark}$) under different excitation intensities at each wavelength is shown in **Figure S5.b**. We observe a linear increase in the magnitude of the ($i_{Tip}/i_{Tip,dark}$) by increasing the excitation intensity at each excitation wavelength. This increasing trend indicates that the local concentration of $Fe(CN)_6^{4-}$ at the tip-substrate gap increases due to an electron-driven reduction reaction at the plasmonic substrate which gets enhanced in kinetics by illumination intensity. We extracted the substrate photocurrent ($i_{sub,photo}$) using the calibration curve (**Figure S5.c**) obtained by a diffusion model[8] and the measured $i_{Tip}/i_{Tip,dark}$ under different illumination intensities (**Figure S5.b**) for each excitation wavelength, and plotted as a function of power in **Figure S5.d**. As a control experiment, we performed the same SECM measurement on a bare p-GaN substrate in the absence of the Au NDs. An extremely low photo-induced ($i_{Tip}$) enhancement was observed at short wavelengths (470 nm-480 nm) by increasing the excitation intensity (**Figure S6**). This was therefore subtracted from the Au/p-GaN photocurrent. The external and internal quantum efficiency (EQE and IQE) of the photocathodes were determined (see **Supplementary Information 1.4**) and plotted in **Figure 2.b,c**, respectively. The magnified view in **Figure 2.b** shows a small EQE peak associated with the characteristic peak plasmon absorption in each structure, before exhibiting a rapid increase beyond 2.4 eV. Conversely, the IQEs (**Figure 2.c**) are largely featureless from 1.4 to 2 eV. They show a minor bump at 2.1 eV and reach their maximum efficiency in the interband regime with the best performance for the thinnest (14 nm thick) NDs device. Considering this is an electron-driven reaction, This trend is surprising because with interband excitation hot electrons in Au are expected to have preferentially low energies[26] that should lead to a decrease in the probability of hot electron tunneling onto the redox molecule compared to lower photon energies.

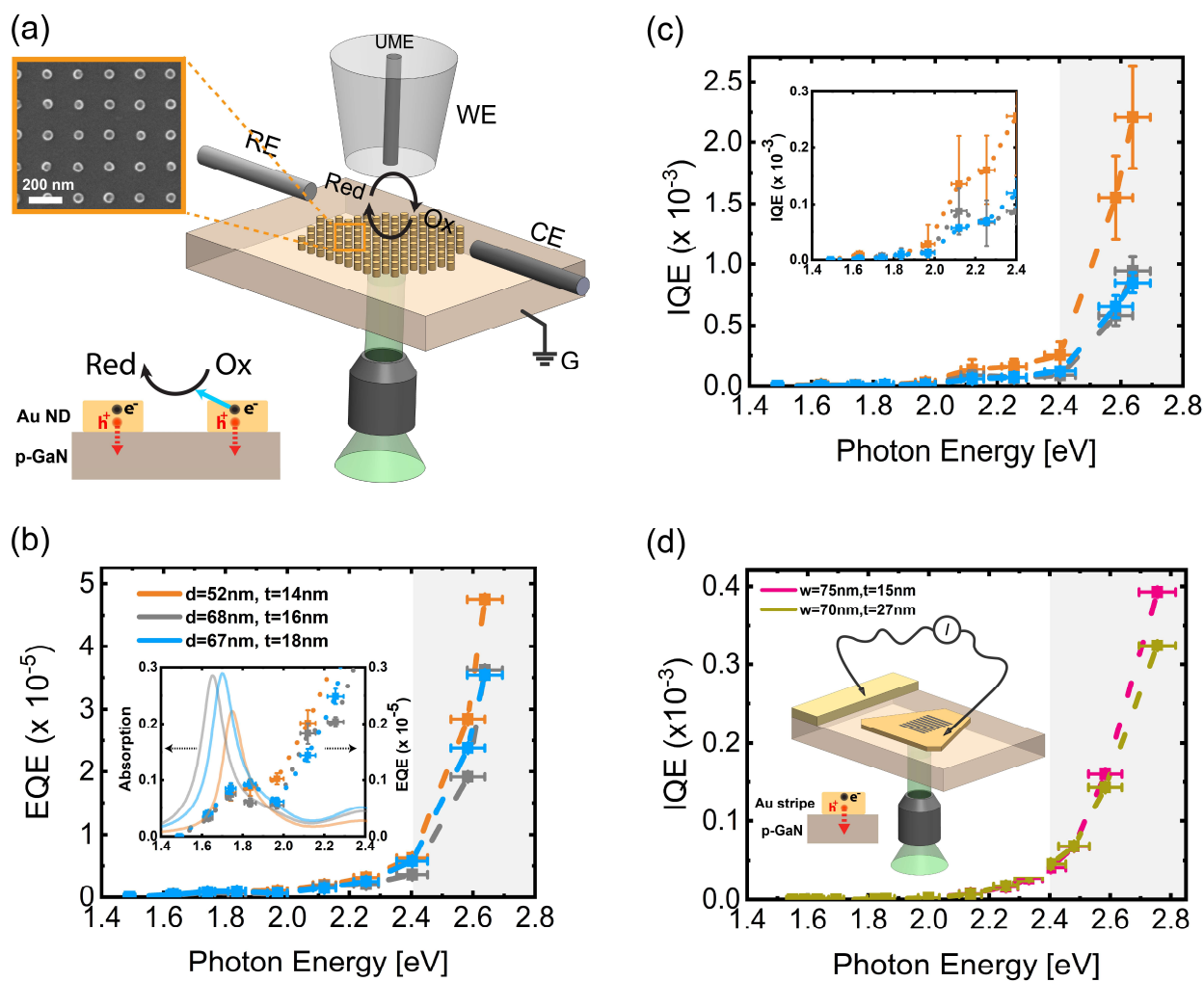

**Figure 2**. Liquid-state photochemical and solid-state photocurrent measurement results. (a) Schematic of the designed plasmonic heterostructure and photo-SECM configuration in an substrate generation/tip collection (SG/TC) experiment mode. A fabricated Au NDs array from a single-crystalline Au microflake on p-GaN substrate is in contact with an electrolyte contacting 4mM $Fe(CN)_6^{3-}$ (oxidant, *Ox*) and 0.25M KCl. A 1.45 μm-radius Pt UME tip is positioned 3 μm away from the substrate. The UME tip is biased at 0.4 V vs Ag/AgCl (reference electrode, *RE*), and the substrate is at open circuit and grounded. Light is incident on the plasmonic Au NDs array from the bottom. The reverse reaction happens at the tip electrode and substrate surface. The current is measured through the tip working electrode (*WE*). A Pt wire is used as a counter electrode (*CE*) to complete the circuit. The inset shows the SEM image of the Au ND array with an average diameter of 52 nm and thickness of 14 nm. The side-view schematic in (a) illustrates the direction of carrier transfer at interfaces. Experimentally determined (b) external quantum efficiency (EQE) and (c) internal quantum efficiency (IQE) spectra for the fabricated heterostructures having different Au ND dimensions. The inset in (b) shows the magnified view of the EQEs together with the calculated absorption spectra and the inset in (c) shows IQEs from 1.4-2.4 eV. (d) Measured IQE spectra of the fabricated heterostructures having thicknesses of 15 and 27 nm. The inset shows the schematic of the designed plasmonic Schottky photodiode heterostructure and solid-state measurement configuration. A stripe Au pattern is fabricated from an SC Au MF on p-GaN substrate together with a 100 nm thick sputtered Au film Ohmic contact. Light is incident on the plasmonic Au stripe array from the bottom and the photocurrent is collected by two microcontact probes electrically connected to the Au flake and the sputtered Au contact pad. The side-view schematic in (d) illustrates the direction of hole transfer at the Au/p-GaN interface. The gray shaded areas depict the purely interband region and the dashed lines are a guide to the eye in panels (b), (c), and (d).

In our photocathode heterostructures, there is a simultaneous collection of hot electrons and hot holes at the metal/electrolyte and metal/semiconductor interfaces, respectively. Therefore, we first briefly focus on the solid-state part to understand the role of the metal/semiconductor interface. For solid-state measurements, we use a plasmonic Schottky photodiode devices consisting of an array of monocrystalline Au stripes with a thickness of 15 and 27 nm on the same p-GaN substrate as the photochemical experiments. This geometry enables easy electrical contacting to the metal while exhibiting plasmonic modes comparable to the disk ones (**Figure S3.b** and **S8**, absorption spectra). We illuminated the substrate through the bottom (**Figure 2.d**, inset) to measure absorption and, subsequently the photocurrent signal. **Figure S7** shows the time-trace of the short-circuit current ($I_{SC}$) upon illumination of the 15 nm thick stripe array (**Figure S4.a**) with wavelengths from 450 to 800 nm at different powers. Subsequently we calculate EQE (**Figure S8**) and IQE (**Figure 2.d**) spectra of each hot hole photodetector. Contrary to the photoelectrochemical device, thicker and thinner nanoantennas exhibit comparable IQE values. Indeed, with bottom illumination the largest hot carrier generation occurs close to the bottom (i.e. metal/semiconductor) interface and, as a result, the non-uniform generation field profile favors ballistic hot hole collection at the Au/p-GaN interface, mitigating the effect of increasing antenna thickness[26] (**Figure 2.d**). We further observe that the steeply increasing portion of the IQEs of the photodiode devices is consistent with theoretical predictions of the energy distributions of photo-excited hot carriers generated via interband transitions.[27] As the incident photon energy is continually increased above the onset of the interband threshold for Au (hv ~> 1.8 eV), an ever-increasing fraction of hot holes are generated within the d-bands of the metal. These hot holes possess enough energy to overcome the interfacial Schottky barrier ($\Phi_B$ =1.3 eV, **Figure S4.b**) and effectively inject into the p-GaN valence band.

When we compare the IQE spectra of the photocathode and photodiode devices, an interesting observation arises: they both exhibit a remarkably similar trend, and the magnitude of the IQE for the photocathodes is surprisingly higher than the IQE of the hot-hole collection in photodiodes. Based on these observations, one may conclude that the IQE of the photocathodes is limited by hot hole collection at the solid-solid interface. To investigate this, we performed the photo-SECM experiment at a concentration of $Fe(CN)_6^{3-}$ that is four times lower (i.e. 1 mM oxidant) in the electrolyte, to see whether the IQE of the photocathode changes with concentration or not. **Figure 3.a** shows the IQE of the 14 nm Au NDs/p-GaN photocathode in 1mM and 4mM $Fe(CN)_6^{3-}$ electrolyte. We clearly observe a change in the IQE magnitude directly proportional to the concentration of oxidant, while preserving the energy dependence (the same IQE trend). This therefore suggests that, contrary to the initial hypothesis, the system is controlled by the hot-electron-driven reaction rather than the hot-hole collection. Indeed, in the latter case, we should not see a concentration dependence in the reaction rate, and thus IQE magnitude. The higher magnitude of the IQE for the photocathode is likely due to a fast and effective electron collection at the metal/molecule interface.

To understand how the electron collection controls the IQE in this system and explain the unexpected trend of IQE with photon energy (**Figure 2.c**), we employed the NESSE transport model[8,28] to calculate the fluxes of hot electrons reaching the metal/electrolyte interface before and after *N* scattering events ($F_N$ *(E, ℏω),* N=0-4 where *E* is the energy of the hot carriers, and *ℏω* is the photon energy). Next, we used a stochastic fitting of the experimental IQE spectra to estimate the injection probability ($P_{inj}(E)$) of the hot electrons to the molecule, respectively (see **Supplementary Information 1.5**). **Figure 3.b** shows the calculated energy-resolved electron fluxes reaching the top surface of 14 nm thick Au NDs directly or upon scattering under illumination at 470 nm (2.64 eV, carriers are generated via interband electron transitions). The initial distribution (N=0) shows the maximum flux for the low-energy hot electrons close to the Fermi level and this portion increases for the cumulative fluxes after scattering events (N=1-4), while the tiny flux of high-energy hot electrons (> 2 eV) remained unchanged. **Figure 3.c** shows the resulting $P_{inj}(E)$ of hot electrons collected from the top {111} and side {110} surfaces[8,21] of the NDs. As the flux distribution of initial and homogenized carriers does not change significantly at energies above 2 eV, no difference was obtained in $P_{inj}(E)$ of high-energy hot electrons (> 2eV) in non-equilibrium (**Figure S9**,

N=0) and steady-state (**Figure 3.c**, N=4) conditions, showing that they can only be collected ballistically. In comparison, we can see a difference in $P_{inj}$ *(E)* for the intermediate energies. Interestingly, regardless of the scattering contribution, our energy-resolved hot electron injection probability (**Figure 3.c**, N=4) shows the coexistence of two distinct hot electron transfer that can be explained with these two mechanisms: (i) a tunneling contribution that exponentially increases with increasing electron energy: this is consistent with an outer-sphere electron transfer and is similar to the $P_{inj}$ *(E)* obtained for outer-sphere hot-hole-driven oxidation of $Fe(CN)_6^{4-}$,[8] and (ii) an efficient transfer of low energy electrons around the lowest unoccupied molecular orbital (LUMO) of the $Fe(CN)_6^{3-}$ molecule, which is about 0.19 eV with respect to the Fermi level of the system (**Figure S2**). The latter contribution can be explained considering the low reduction potential or LUMO level of $Fe(CN)_6^{3-}$ and a different mechanism for electron transfer to the LUMO than an outer-sphere transfer model observed for the hole injection to the HOMO of $Fe(CN)_6^{4-}$.[8] Indeed, surface-enhanced Raman spectroscopy (SERS) studies[17,24,25,29] have demonstrated that electron transfer in the ferri-/ferrocyanide system cannot be strictly considered as a purely outer-sphere process. In $Fe(CN)_6^{3-}$ molecule, the interaction between the Fe d-orbitals and the π*-antibonding orbitals of the cyanide (CN) ligands results in the formation of molecular orbitals. The HOMO primarily originates from the Fe d-orbitals and is associated with electron density around the Fe atom. The LUMO corresponds mainly to the π*-antibonding orbital of the CN ligands, which exists in a region of lower electron density surrounding the CN ligands and exhibits a strong affinity for accepting electrons.[30] SERS studies have indicated the presence of bridging CN ligands in $Fe(CN)_6^{3-}$ molecules adsorbed on the Au surface.[17,24] These ligands form bonds, involving at least one CN group, with the Au surface through the lone-pair electrons (σ) on the N atoms. These interactions mainly occur in a CN antibonding orbital, enabling efficient electron transfer from Au to the LUMO of the $Fe(CN)_6^{3-}$ molecule. Moreover, it has been shown that the $Fe(CN)_6^{4-}$ is not strongly adsorbed on the Au surface, as the Fe-C≡N bonding is stronger than the C≡N bond, whereas the opposite is true for $Fe(CN)_6^{3-}$, where the oxidation state of Fe increased.[24,29]

Therefore, the efficient inner-sphere transfer of low-energy electrons to the LUMO (≈0.19eV) results in the observed efficient $P_{inj}$ *(E)* (**Figure 3.c**). Combined with the large flux of low-energy hot electrons (**Figure 3.b**, blue area), significantly higher compared to the high-energy ones (**Figure 3.b**, gray area), this results in the rapid IQE growth in the interband regime (**Figure 2.c**, gray area). If these low-energy electrons are not collected, a dramatic drop in IQE should be observed in the interband regime due to the decrease in the number of high-energy electrons, similar to the IQE of electron-collection $Au/TiO_2$ photodetector device.[8]

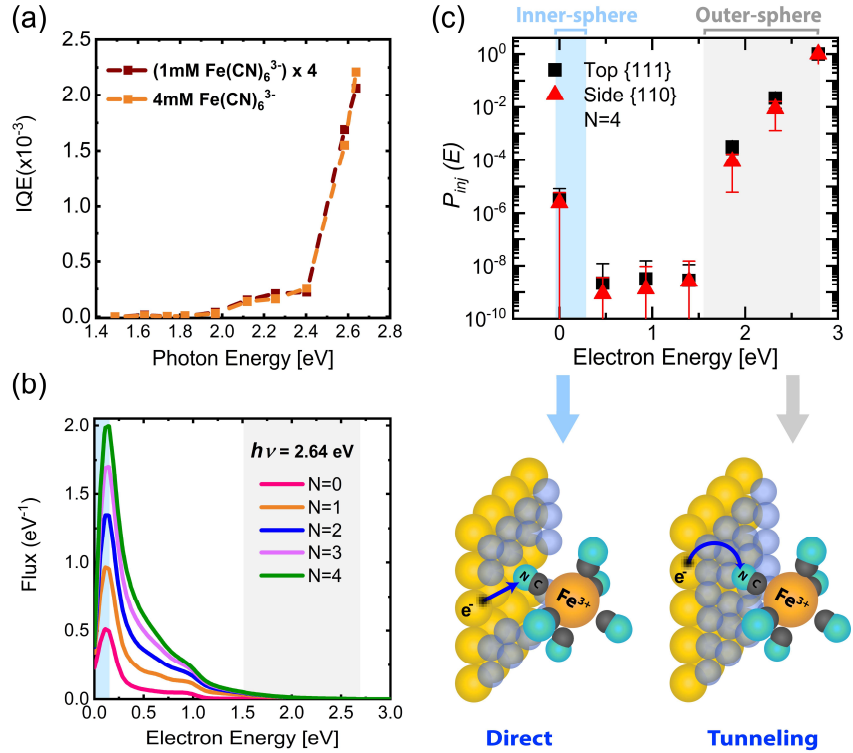

**Figure 3.** Hot electron generation, transport, and injection in Au ND/p-GaN photocathode devices. (a) IQE of the 14 nm Au NDs/p-GaN photocathode in 1mM (x4) and 4mM $Fe(CN)_6^{3-}$ electrolyte concentrations. The IQE magnitude is directly proportional to the concentration of $Fe(CN)_6^{3-}$, while the energy dependence remained conserved. The dashed lines are a guide to the eye. (b) Calculated energy resolved electron fluxes reaching the top surface of 14 nm thick Au NDs directly (N=0) or upon scattering (N=1-4) under illumination at 470 nm (2.64 eV). The blue shaded area shows the position of the LUMO with respect to the Fermi level (0.19 eV). The gray shaded area shows the tiny flux of high-energy hot electrons (1.5-2.64 eV). (c) Injection probability ($P_{inj}(E)$) for hot electrons collected from the top {111} and side {110} facets of the Au NDs after 4 scattering events. The $P_{inj}(E)$ plots are extracted from the fitting approach using the energy-resolved hot electron fluxes and experimentally determined IQEs of Au NDs heterostructures. The direct (inner-sphere) and tunneling (outer-sphere) charge injection mechanism is indicated for the low-energy and high-energy regions, respectively.

**Figure 3.c** also shows a comparable $P_{inj}(E)$ for the hot electrons collected from the top {111} and side {110} facets of the Au NDs. This can be related to the comparatively long mean free path of hot electrons (10 nm for a 2eV hot electron),[27] which allows them to be collected more efficiently from both surfaces

compared to the hot holes (see our previous work).[8] Moreover, all the ND structures studied here are ultrathin (with thicknesses < 20 nm), resulting in a more uniform generation profile of hot carriers across the volume of the NDs.[8] Our transport model and fitting approach for injection probability reproduced our experimental data well, and almost the same computed IQE trend was obtained up to 2.5 eV (**Figure S10**). The higher discrepancy for the higher-energy values is likely due to the very small flux of high-energy electrons obtained within this range of incident photon energy. An analogous analysis applied to the hot holes in the Au stripe/p-GaN photodetector system is reported in **Supplementary Information 5**, **Figure S11**. The results indicate a ballistic collection of the high-energy nonequilibrium hot holes at the Au/p-GaN interface.

**Conclusion**

In summary, we simultaneously investigated the transport and collection of hot carriers in ultrathin (14-18 nm) single-crystalline plasmonic Au nanoantenna array Schottky photodiode and photocathode devices that operate via the collection of hot holes (Au/p-GaN) and both hot holes and electrons ($Fe(CN)_6^{3-}$/Au/p-GaN), respectively. Our internal quantum efficiency analysis at different redox concentrations demonstrated the performance of the photocathode device in which both hot electrons and hot holes are collected, is controlled by the hot-electron collection at the metal/molecule interface. Furthermore, our injection probability model for this hot-electron-driven reduction reaction revealed a charge transfer mechanism that deviates from the purely outer-sphere mechanism observed for the oxidation of $Fe(CN)_6^{4-}$.[8] This observed mechanism showed two contributions: low-energy electrons transferring directly from Au to the LUMO of the $Fe(CN)_6^{3-}$ molecule via an inner-sphere process and higher-energy electrons transferring to the molecule through an outer-sphere tunneling process. The efficient collection of lower-energy electrons, given their higher flux compared to higher-energy ones, results in a continuous increase in the IQE of the photocathode device in the interband regime. Additionally, our solid-state study showed a ballistic collection of high-energy d-band holes at the Au/p-GaN interface. At this stage, our results shed light on important mechanisms governing the transport and injection of hot carriers across interfaces in hot-carrier-driven photocatalytic systems. Our methodology can be a powerful tool to guide the design of efficient hot-carrier-driven devices, particularly plasmon-driven artificial photosynthetic systems. Future experiments with surface-enhanced Raman spectroscopy (SERS) may expand our understanding by identifying intermediate chemical species adsorbed on the surface, providing detailed insights into reaction pathways, especially for complex, multiproduct reactions involving inner-sphere charge transfer mechanisms.

**Supporting Information**

Methods section; additional information about characterization of the redox molecule/Au/p-GaN system; characterization of the Au nanoantenna heterostructures; substrate photocurrent response in photo-SECM and solid-state measurements; detailed analysis of IQE, transport and injection probability results.

**Acknlowledgements**

The authors acknowledge the support of the Swiss National Science Foundation (Eccellenza Grant #194181). ARB acknowledges SNSF Swiss Postdoctoral Fellowship TMPFP2_217040. The authors also acknowledge the support of the following experimental facilities at EPFL: Center of MicroNanoTechnology (CMi), and Interdisciplinary Centre for Electron Microscopy (CIME). The GaN wafers were provided by the Advanced Semiconductors for Photonics and Electronics (LASPE) group at EPFL. The Calculations (using the NESSE framework) were performed at the Center for Computational Innovations at Rensselaer Polytechnic Institute.

**Author Information**

Corresponding Author

Giulia Tagliabue, Ecole Polytechnique Federale de Lausanne (EPFL), Lausanne, Switzerland, orcid.org/0000-0003-4587-728X, Email: giulia.tagliabue@epfl.ch

Other Authors

Fatemeh Kiani, Ecole Polytechnique Federale de Lausanne (EPFL), Lausanne , Switzerland, orcid.org/0000-0002-2707-5251

Alan R. Bowman, Ecole Polytechnique Federale de Lausanne (EPFL) , Lausanne, Switzerland, orcid.org/0000-0002-1726-3064

Milad Sabzehparvar, Ecole Polytechnique Federale de Lausanne (EPFL) , Lausanne , Switzerland, orcid.org/0000-0001-5594-6889

Ravishankar Sundararaman, Rensselaer Polytechnic Institute, New York, USA, orcid.org/0000-0002-0625-4592

**References**

(1) Song, K.; Lee, H.; Lee, M.; Park, J. Y. Plasmonic Hot Hole-Driven Water Splitting on Au Nanoprisms/P-Type GaN. *ACS Energy Lett.* **2021**, *6* (4), 1333–1339.

(2) DuChene, J. S.; Tagliabue, G.; Welch, A. J.; Li, X.; Cheng, W.-H.; Atwater, H. A. Optical Excitation of a Nanoparticle Cu/p-NiO Photocathode Improves Reaction Selectivity for CO2 Reduction in Aqueous Electrolytes. *Nano Lett.* **2020**, *20* (4), 2348–2358.

(3) Li, R.; Cheng, W.-H.; Richter, M. H.; DuChene, J. S.; Tian, W.; Li, C.; Atwater, H. A. Unassisted Highly Selective Gas-Phase CO2 Reduction with a Plasmonic Au/p-GaN Photocatalyst Using H2O as an Electron Donor. *ACS Energy Lett.* **2021**, *6* (5), 1849–1856.

(4) DuChene, J. S.; Tagliabue, G.; Welch, A. J.; Cheng, W.-H.; Atwater, H. A. Hot Hole Collection and Photoelectrochemical CO2 Reduction with Plasmonic Au/p-GaN Photocathodes. *Nano Lett.* **2018**, *18* (4), 2545–2550.

(5) Cortés, E. Efficiency and Bond Selectivity in Plasmon-Induced Photochemistry. *Adv. Opt. Mater.* **2017**, *5* (15), 1700191.

(6) Kazuma, E.; Jung, J.; Ueba, H.; Trenary, M.; Kim, Y. Real-Space and Real-Time Observation of a Plasmon-Induced Chemical Reaction of a Single Molecule. *Science* **2018**, *360* (6388), 521–526.

(7) Yu, Y.; Williams, J. D.; Willets, K. A. Quantifying Photothermal Heating at Plasmonic Nanoparticles by Scanning Electrochemical Microscopy. *Faraday Discuss.* **2018**, *210* (0), 29–39.

(8) Kiani, F.; Bowman, A. R.; Sabzehparvar, M.; Karaman, C. O.; Sundararaman, R.; Tagliabue, G. Transport and Interfacial Injection of D-Band Hot Holes Control Plasmonic Chemistry. *ACS Energy Lett.* **2023**, 4242–4250.

(9) Zhou, L.; Swearer, D. F.; Zhang, C.; Robatjazi, H.; Zhao, H.; Henderson, L.; Dong, L.; Christopher, P.; Carter, E. A.; Nordlander, P.; Halas, N. J. Quantifying Hot Carrier and Thermal Contributions in Plasmonic Photocatalysis. *Science* **2018**, *362* (6410), 69–72.

(10) Brongersma, M. L.; Halas, N. J.; Nordlander, P. Plasmon-Induced Hot Carrier Science and Technology. *Nat. Nanotechnol.* **2015**, *10* (1), 25–34.

(11) Boerigter, C.; Campana, R.; Morabito, M.; Linic, S. Evidence and Implications of Direct Charge Excitation as the Dominant Mechanism in Plasmon-Mediated Photocatalysis. *Nat. Commun.* **2016**, *7* (1), 10545.

(12) Linic, S.; Aslam, U.; Boerigter, C.; Morabito, M. Photochemical Transformations on Plasmonic Metal Nanoparticles. *Nat. Mater.* **2015**, *14* (6), 567–576.

(13) Tagliabue, G.; Jermyn, A. S.; Sundararaman, R.; Welch, A. J.; DuChene, J. S.; Pala, R.; Davoyan, A. R.; Narang, P.; Atwater, H. A. Quantifying the Role of Surface Plasmon Excitation and Hot Carrier Transport in Plasmonic Devices. *Nat. Commun.* **2018**, *9* (1), 1–8.

(14) Tagliabue, G.; DuChene, J. S.; Habib, A.; Sundararaman, R.; Atwater, H. A. Hot-Hole versus Hot-Electron Transport at Cu/GaN Heterojunction Interfaces. *ACS Nano* **2020**, *14* (5), 5788–5797.

(15) Tanimoto, S.; Ichimura, A. Discrimination of Inner- and Outer-Sphere Electrode Reactions by Cyclic Voltammetry Experiments. *J. Chem. Educ.* **2013**, *90* (6), 778–781.

(16) Allen J. Bard; Larry R. Faulkner; Henry S. White. *Electrochemical Methods: Fundamentals and Applications*, 3rd ed.; Wiley, 2022.

(17) Qi, Y.; Brasiliense, V.; Ueltschi, T. W.; Park, J. E.; Wasielewski, M. R.; Schatz, G. C.; Van Duyne, R. P. Plasmon-Driven Chemistry in Ferri-/Ferrocyanide Gold Nanoparticle Oligomers: A SERS Study. *J. Am. Chem. Soc.* **2020**, *142* (30), 13120–13129.

(18) Schlather, A. E.; Manjavacas, A.; Lauchner, A.; Marangoni, V. S.; DeSantis, C. J.; Nordlander, P.; Halas, N. J. Hot Hole Photoelectrochemistry on Au@SiO2@Au Nanoparticles. *J. Phys. Chem. Lett.* **2017**, *8* (9), 2060–2067.

(19) Kazuma, E.; Lee, M.; Jung, J.; Trenary, M.; Kim, Y. Single-Molecule Study of a Plasmon-Induced Reaction for a Strongly Chemisorbed Molecule. *Angew. Chem. Int. Ed.* **2020**, *59* (20), 7960–7966.

(20) Kale, M. J.; Avanesian, T.; Xin, H.; Yan, J.; Christopher, P. Controlling Catalytic Selectivity on Metal Nanoparticles by Direct Photoexcitation of Adsorbate–Metal Bonds. *Nano Lett.* **2014**, *14* (9), 5405–5412.

(21) Kiani, F.; Tagliabue, G. High Aspect Ratio Au Microflakes via Gap-Assisted Synthesis. *Chem. Mater.* **2022**, *34* (3), 1278–1288.

(22) Beach, J. D.; Collins, R. T.; Turner, J. A. Band-Edge Potentials of n-Type and p-Type GaN. *J. Electrochem. Soc.* **2003**, *150* (7), A899.

(23) Yu, Y.; Wijesekara, K. D.; Xi, X.; Willets, K. A. Quantifying Wavelength-Dependent Plasmonic Hot Carrier Energy Distributions at Metal/Semiconductor Interfaces. *ACS Nano* **2019**, *13* (3), 3629–3637.

(24) Lowry, R. B. SERS and Fourier Transform SERS Studies of the Hexacyanoferrate(III)-Hexacyanoferrate(II) Couple on Gold Electrode Surfaces. *J. Raman Spectrosc.* **1991**, *22* (12), 805–809.

(25) Loo, B. H.; Lee, Y. G.; Liang, E. J.; Kiefer, W. Surface-Enhanced Raman Scattering from Ferrocyanide and Ferricyanide Ions Adsorbed on Silver and Copper Colloids. *Chem. Phys. Lett.* **1998**, *297* (1), 83–89.

(26) Tagliabue, G.; Jermyn, A. S.; Sundararaman, R.; Welch, A. J.; DuChene, J. S.; Pala, R.; Davoyan, A. R.; Narang, P.; Atwater, H. A. Quantifying the Role of Surface Plasmon Excitation and Hot Carrier Transport in Plasmonic Devices. *Nat. Commun.* **2018**, *9* (1), 3394.

(27) Brown, A. M.; Sundararaman, R.; Narang, P.; Goddard, W. A. I.; Atwater, H. A. Nonradiative Plasmon Decay and Hot Carrier Dynamics: Effects of Phonons, Surfaces, and Geometry. *ACS Nano* **2016**, *10* (1), 957–966.

(28) Jermyn, A. S.; Tagliabue, G.; Atwater, H. A.; Goddard, W. A.; Narang, P.; Sundararaman, R. Transport of Hot Carriers in Plasmonic Nanostructures. *Phys. Rev. Mater.* **2019**, *3* (7), 075201.

(29) Hanusa, T. P. Cyanide Complexes of the Transition Metals. In *Encyclopedia of Inorganic and Bioinorganic Chemistry*; John Wiley & Sons, Ltd, **2011**.

(30) Cotton, F. A.; Wilkinson, G.; Murillo, C. A.; Bochmann, M. *Advanced Inorganic Chemistry*; John Wiley & Sons, **1999**.

**Table of Contents Graphic**

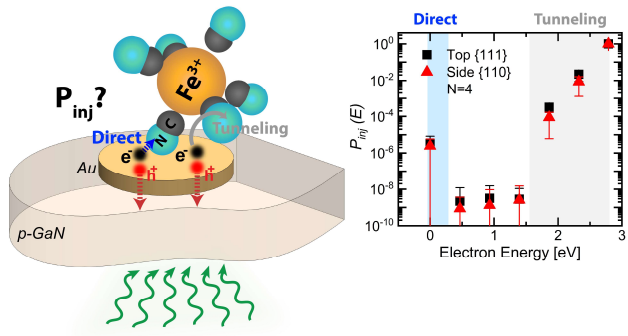